# USE OF COMBINED SCALING OF REAL SEISMIC RECORDS TO OBTAIN CODE-COMPLIANT SETS OF ACCELEROGRAMS: APPLICATION FOR THE CITY OF BUCHAREST

*Iolanda-Gabriela CRAIFALEANU[1,2], Ioan Sorin BORCIA[1]*
[1] NIRD "URBAN-INCERC", INCERC Bucharest Branch, Romania, isborcia@yahoo.com
[2] Technical University of Civil Engineering Bucharest, Romania, i.craifaleanu@gmail.com

**ABSTRACT**

A recently proposed method for scaling real accelerograms to obtain sets of code-compliant records is assessed. The method, which uses combined time and amplitude scaling, corroborated with an imposed value of an instrumental, Arias-type intensity, allows the generation of sets of accelerograms for which the values of the mean response spectrum for a given period range are not less than 90% of the elastic response spectrum specified by the code. The method, which is compliant with both for the Romanian seismic code, P100-1/2006, and Eurocode 8, was described in previous papers. Based on dynamic analyses of single-degree-of freedom (SDOF) and of multi-degree-of-freedom (MDOF) systems, a detailed application and assessment of the method is performed, for the case of the long corner period design spectrum in Bucharest. Conclusions are drawn on the advantages of the method, as well as on its potential improvement in the future.

**REZUMAT**

În articol este evaluată o metodă propusă recent pentru scalarea accelerogramelor reale în scopul obţinerii de înregistrări corespunzătoare cerinţelor codului seismic. Metoda, care utilizează scalarea combinată în domeniul timpului şi al amplitudinii, coroborată cu o valoare impusă a unei intensităţi instrumentale de tip Arias, permite generarea de seturi de accelerograme pentru care valorile spectrului mediu pentru un domeniu de perioade dat nu se situează sub 90% din spectrul de răspuns elastic specificat de cod. Metoda, compatibilă atât cu cerinţele codului seismic românesc, P100-1/2006, cât şi cu cele ale Eurocodului 8, a fost descrisă în articole anterioare. Este efectuată o evaluare detaliată a metodei, pentru cazul spectrului de proiectare cu perioadă de colţ lungă, corespunzător municipiului Bucureşti, pe baza analizelor dinamice efectuate asupra sistemelor cu un grad, respectiv cu mai multe grade de libertate dinamică. Sunt obţinute concluzii asupra avantajelor metodei, ca şi asupra potenţialelor sale perfecţionări viitoare.

*Cuvinte cheie*: cod seismic; scalarea accelerogramelor; intensitatea Arias; selecţia înregistrărilor seismice; aproximarea spectrului

## 1. INTRODUCTION

The dynamic analysis of structures according to seismic code regulations requires the selection / generation of sets of accelerograms complying with certain relevance criteria. If real (recorded) accelerograms are used, these criteria concern the adequacy to the seismogenetic features of the sources and to the soil conditions appropriate to the site, as well as the scaling of their values to the appropriate peak ground acceleration, as specified by the code for the zone under consideration. Additionally, there are requirements concerning the maximum allowed differences between the elastic acceleration response spectrum provided by the spectrum and the mean spectrum calculated for all accelerograms in the set.

For instance, Eurocode 8 (CEN, 2004) requires that "in the range of periods between $0.2T_1$ and $2T_1$, where $T_1$ is the fundamental period of the structure in the direction where the accelerogram will be applied; no value of the mean 5% damping elastic spectrum, calculated from all time histories, should be less than 90% of the corresponding value of the 5% damping elastic response spectrum".

The Romanian seismic code, P100-1/2006 (MTCT, 2006) includes a similar requirement, without, however, explicitly specifying the period range.

A method was proposed recently for obtaining sets of accelerograms compliant with this requirement (Borcia, 2010, Borcia and Dobre, 2012). The method uses the combined time and amplitude scaling of real accelerograms, selected from a record database. Scaling is made such that all accelerograms in the set have the same value of an Arias-type instrumental intensity.

Based on this method, a set of ten accelerograms was generated for the present study, with a mean response spectra compatible with the elastic acceleration response spectrum specified for Bucharest by the Romanian seismic code. This spectrum, shown in Fig. 1, is characterized by a long corner (control) period, $T_C$ = 1.6 s, which takes into account the soft soil conditions of the city and of its surrounding zone.

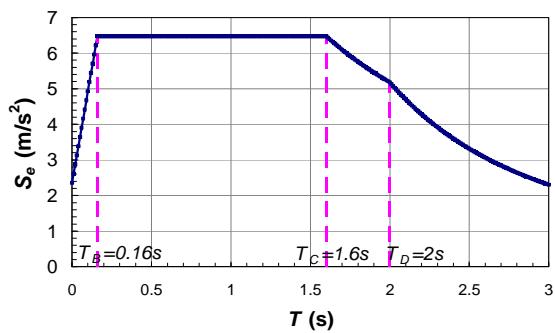

**Fig. 1.** P100-1/2006: elastic acceleration response spectrum for Bucharest (corner period: $T_C$ = 1.6 s, design peak ground acceleration: $a_g$ = 0.24 g)

The effectiveness of the method was assessed by detailed analyses performed on single-degree-of-freedom (SDOF) and multi-degree-of-freedom (MDOF) systems.

## 2. DESCRIPTION OF THE ACCELEROGRAM SCALING METHOD

In the proposed method, the accelerograms are first selected according to their spectral contents, which should be as closer as possible to that reflected by the target spectrum. Only horizontal components are considered.

Then, the accelerograms are scaled in the time range (which also represents a scaling in the period range), in order to obtain accelerograms with the maximum amplifications in the desired period range.

The resulting accelerograms are further scaled in the amplitude range, so that, for all of them, the instrumental Arias-type intensity, $I_A$, preserves its initial value and the corresponding peak ground acceleration (PGA) is at least equal to the design peak ground acceleration, $a_g$, provided by the P100-1/2006 code for the considered site.

The Arias-type instrumental intensity used above as a scaling criterion is given by the following expression (Sandi et al., 2010, Sandi and Borcia, 2011):

$$I_A = \log_{7.5} \int [w_g(t)]^2 dt + 7.14 \qquad (1)$$

where $w_g(t)$ is the ground acceleration. Proposed by Horea Sandi (Sandi, 1987, Sandi et al., 1998), this intensity is calibrated for compatibility with the EMS macroseismic scale.

## 3. APPLICATION FOR THE CITY OF BUCHAREST

### 3.1. Generation of the accelerogram set

Based on the analysis of available accelerograms in the database of strong motion records compiled by INCERC, three complete ground motion records obtained in Bucharest were chosen. Of those, one is from the March 4, 1977 earthquake (moment magnitude $M_w$ = 7.4) and the other two are records from the August 30, 1986 earthquake ($M_w$ = 7.1).

It should be noted that the first record, obtained at INCERC Bucharest (codified "771inc") is the only available from the 1977 seismic event and that the information it provided was essential in establishing the shape of the code spectrum for Bucharest.

The 1986 records were obtained at the seismic stations Magurele ("86mag") and

I. S. Borcia, I. G. Craifaleanu

EREN ("86exp"), one located in the southwest, and the other in the northwest of Bucharest.

The horizontal components of the three records were scaled in amplitude in order to reach $I_A = 8.4$, with a corresponding PGA of at least 0.24 $g$ (as required for Bucharest by the P100-1/2006 code). The resulting accelerograms were denoted by "77incl", "77inct", "86magl", "86magt", "86expl" and "86expt", where the last letter of the record code identifies the longitudinal and transversal components, respectively.

To obtain a set of accelerograms with a mean spectrum that approximates the code spectrum as required by the P100-1/2006 code, the "771inc" records were further scaled in time, then in amplitude. Thus were obtained the accelerograms "771p6inc" (l & t), with the maximum spectral amplification at $T = 1.6$ s, and "771pinc" (l & t), with the maximum spectral amplification at $T = 1.0$ s.

The characteristic parameters of the resulting set of 10 accelerograms are shown in Table 1.

**Table 1.** *Characteristic parameters of the considered accelerogram set*

|  | 86expl | 86expt | 86magl | 86magt | 77incl | 77inct | 771pincl | 771pinct | 771p6incl | 771p6inct |
|---|---|---|---|---|---|---|---|---|---|---|
| $\Delta t$ (s) | 0.010 | 0.010 | 0.010 | 0.010 | 0.005 | 0.005 | 0.00411 | 0.00411 | 0.00658 | 0.00658 |
| PGA (m/s²) | 4.97 | 3.82 | 3.80 | 3.62 | 3.53 | 3.13 | 3.90 | 3.45 | 3.08 | 2.72 |
| PGA initial (m/s²) | 1.61 | 1.06 | 1.35 | 1.15 | 1.88 | 2.07 | - | - | - | - |
| $I_A$ initial | 7.28 | 7.13 | 7.38 | 7.20 | 7.77 | 7.99 | - | - | - | - |
| $I_A$ (PGA=0.24$g$) | 7.66 | 7.92 | 7.93 | 8.01 | 8.00 | 8.12 | - | - | - | - |
| $I_A$ final | 8.4 | | | | | | | | | |

## 3.2. Linear and nonlinear response spectra

### 3.2.1. Acceleration spectra

Linear and nonlinear acceleration spectra were computed for the considered set of accelerograms, in order to assess the effectiveness of the scaling method.

As a parameter of spectral curves, the strength reduction factor, $R$, was chosen:

$$R = F_{el,\max} / F_y \qquad (2)$$

where $F_{el,max}$ is the force induced by the seismic motion to the system, in the hypothesis of unlimited elastic behavior, and $F_y$ is the yield force of the nonlinear system. The value $R = 1$ corresponds to linear behavior.

Spectra were computed for SDOF systems with a damping factor of 5%, by considering an elastic-perfectly plastic hysteretic rule and $R$ values from 1 to 10, with a step of one.

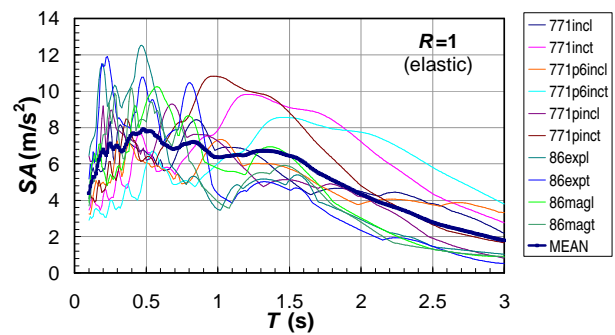

**Fig. 2.** Linear acceleration response spectra for the considered accelerogram set ($R = 1$)

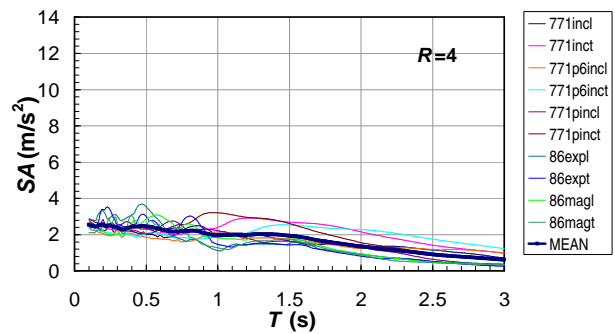

**Fig. 3.** Nonlinear acceleration response spectra for the considered accelerogram set ($R = 4$)

Fig. 2 shows linear acceleration spectra computed for the whole set of accelerograms, together with the mean spectrum, while in Fig. 3 nonlinear acceleration response spectra for the same set and $R = 4$ are displayed.

Design acceleration spectra for Bucharest were computed, for comparison, according to the P100-1/2006 code, for behavior factors ranging from $q = 1$ to 10. It should be noted that the definition of the behavior factor, $q$, in the Romanian seismic code is similar to that in Eurocode 8, but its values are generally higher (Craifaleanu, 2008).

Mean acceleration spectra for the considered accelerogram set, together with design spectra, are shown in Fig. 4.

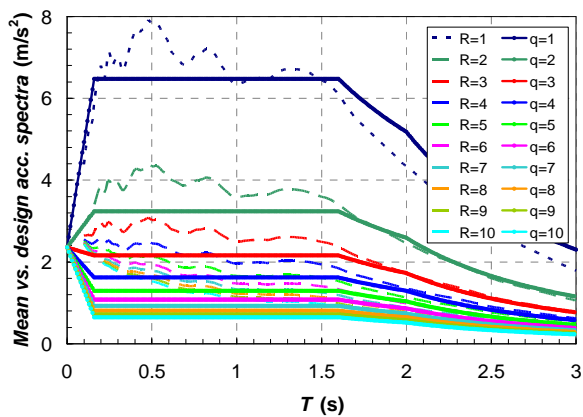

**Fig. 4.** Mean acceleration spectra for the considered record set vs. design spectra specified for Bucharest by the P100-1/2006 code

Fig. 5 shows the ratio between mean acceleration spectra and the corresponding design spectra.

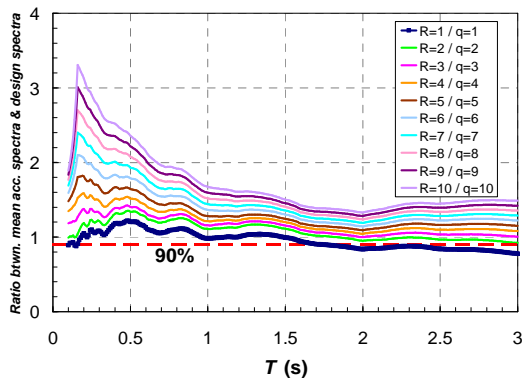

**Fig. 4.** Ratio between mean acceleration spectra for the considered record set and corresponding design spectra for Bucharest

As it can be observed from the above graphs, the mean linear acceleration spectrum follows rather closely the elastic spectrum specified by the P100-1/2006 code, while the "90%" rule is respected practically for the whole range of periods up to 1.77 s, which is the expected period range for most of the usual buildings.

In what concerns nonlinear spectra, mean values obtained for the considered accelerogram set are, without exception, larger than the corresponding design spectra. The largest difference, which appears for $T = T_B = 0.16$ s, is due to the abrupt change of the shape of the design spectrum at that point, which is difficult to follow in the proposed method. Moreover, it should be mentioned that a better concordance would be achieved for $T_B = 0.32$ s, the value proposed by the new edition of the Romanian seismic code, planned to be enforced in 2013.

For larger periods, the differences decrease considerably. It can be observed that, for moderate values of the strength reduction factor (e.g. $R \leq 7$), the ratio in Fig. 5 drops below 2 at periods larger than 0.3 s. These differences are on the safe side from the design point of view.

### 3.2.2. Displacement spectra

Linear and nonlinear displacement spectra were computed for the considered accelerogram set in the same hypotheses as those used for acceleration spectra.

Fig. 5 shows linear displacement spectra computed for the whole set of accelerograms, together with the mean spectrum, while in Fig. 6 nonlinear displacement response spectra for the same set and $R = 4$ are displayed.

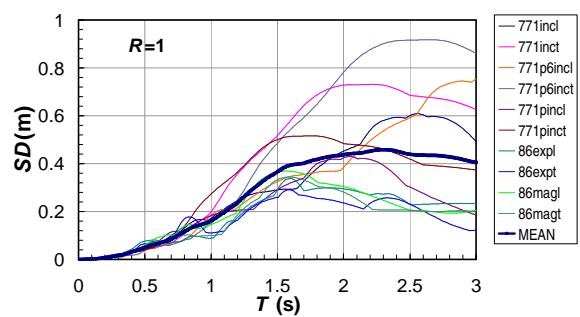

**Fig. 5.** Linear displacement response spectra for the considered accelerogram set ($R = 1$)

I. S. Borcia, I. G. Craifaleanu

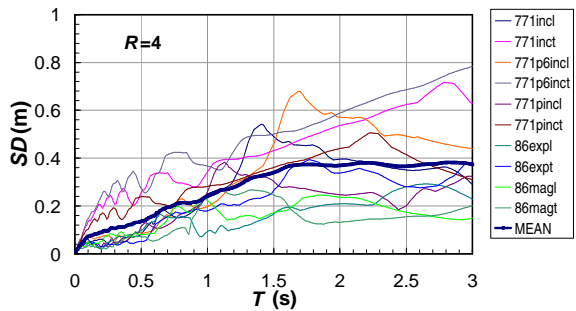

**Fig. 6.** Nonlinear displacement response spectra for the considered accelerogram set ($R = 4$)

Design displacement spectra were computed for Bucharest, according to the provisions of the P100-1/2006 code and of Eurocode 8. Linear spectral displacements were computed by multiplying corresponding spectral accelerations with $T^2/(4\pi^2)$, as in equations (3.7) from both codes. In order to determine nonlinear spectral displacements, the resulting linear displacements were multiplied with displacement amplification coefficients, $c$, as specified by the two codes.

In P100-1/2006, $c$, is given by formula (E.3) in Annex E of the code:

$$1 \leq c = 3 - 2.5 \frac{T}{T_C} \leq 2 \quad (3)$$

where $T_C$ is the corner (control) period of the code spectrum. The above formula is applicable for $q \geq 2$. As one can observe, the displacement amplification factor does not depend on the behavior factor.

For the analysis according to Eurocode 8, the equivalent of $c$ was taken from formula (B.10) in Annex B of the norm. With the required adaptation for SDOF systems, it results

$$1 \leq c = \frac{1}{q}\left[1 + (q-1)\frac{T_C}{T}\right] \leq 3 \quad (4)$$

Thus, the displacement amplification factor specified by the European norm is a function of the behavior factor, $q$, the corner (control) period, $T_C$, and the vibration period, $T$.

The resulting displacement spectra are shown in Figs. 7 and 8, for P100-1/2006 and Eurocode 8, respectively, together with the mean displacement spectra computed for the considered accelerogram set.

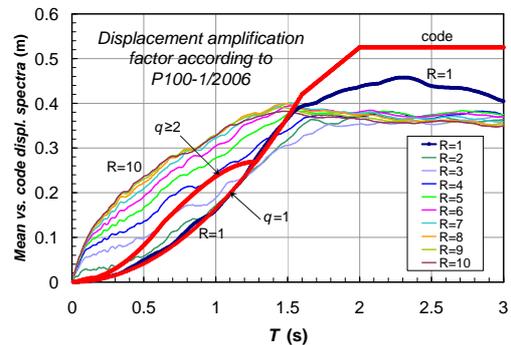

**Fig. 7.** Mean displacement spectra for the considered accelerogram set vs. code displacement spectra (displacement amplification factor computed according to P100-1/2006)

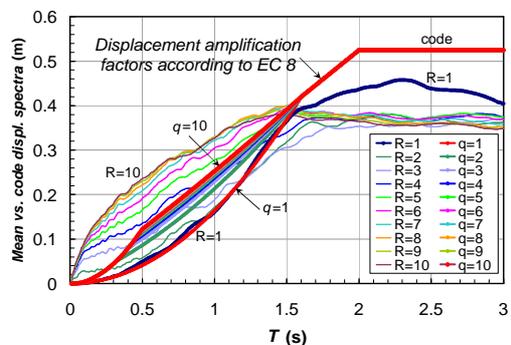

**Fig. 8.** Mean displacement spectra for the considered record set vs. code displacement spectra (displacement amplification factor computed according to Eurocode 8)

For the period range below $T_C = 1.6$ s, a rather good match can be observed, in the above figures, between the mean linear displacement spectrum ($R = 1$) for the considered accelerogram set and the code displacement spectrum ($q = 1$). In what concerns the nonlinear displacement, the match is better when the displacement amplification factor in Eurocode 8 is used, due to the consideration of the variation of $c$ with the behavior factor, $q$, and the corner period, $T_C$.

### 3.3. Results obtained from the analysis of MDOF systems

#### 3.3.1. Methodology and hypotheses

The ten accelerograms of the set were used as input for linear dynamic analyses performed on a simple, 9-storey frame structure. The model of the structure, obtained with SAP2000 (CSI, 2009), is shown in Fig. 9.

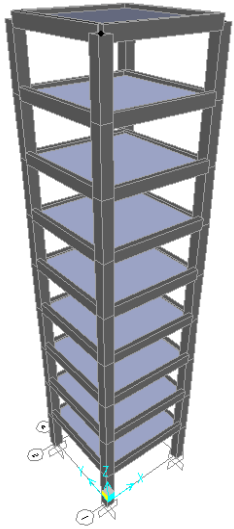

**Fig. 9.** Model of the simple MDOF system used in dynamic analyses

The structure is symmetric about both X and Y axes; the plan dimensions are 6 × 6 m and the story height is 3 m. The properties of the model were calibrated such as to obtain a fundamental vibration period $T_1 = 1$ s.

As a basis for comparison, the seismic response of the structure was computed by using the modal response spectrum analysis, according to the Romanian code. This method is quite similar in both considered codes, P100-1/2006 and Eurocode 8. Seismic actions considered in design were established according to the specifications of the Romanian code.

*3.3.2. Results*

The results were expressed in terms of ratios between maximum values computed by linear dynamic time-history analysis and by the modal response spectrum analysis method (linear case, $q = 1$), respectively. Base shear ratios are shown in Fig. 10, while roof displacement ratios are displayed in Fig. 11.

As it can be observed, the resulting mean ratios, for the entire set, are practically equal to one. This indicates that using the considered set of accelerograms as an input for linear analysis leads to a seismic response that is very close to the response under code-specified actions.

The result confirms the positive conclusions obtained in the previous section by the analysis of response spectra and shows the efficacy of the proposed method in generating code-compliant sets of accelerograms.

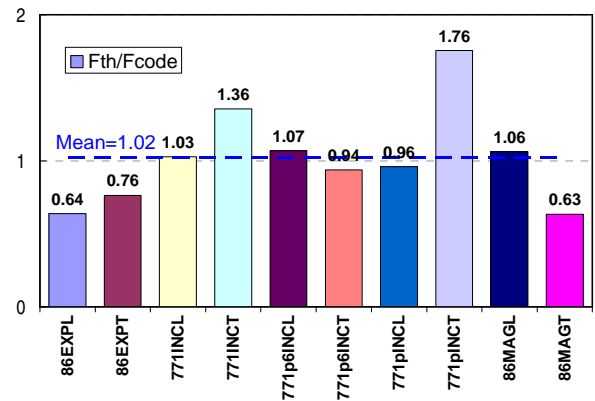

**Fig. 10.** Ratios between maximum base shears determined by linear dynamic analysis ($F_{th}$) and by modal response spectrum method ($F_{code}$)

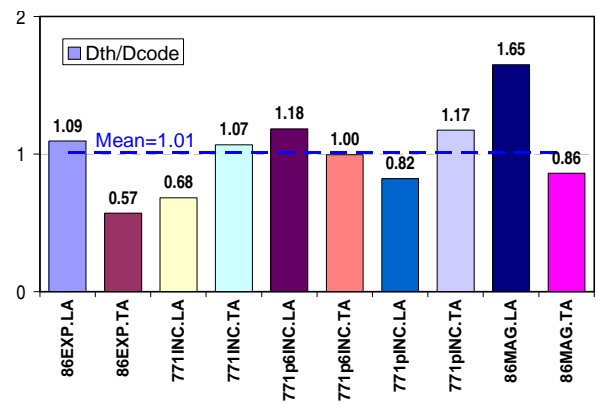

**Fig. 11.** Ratios between maximum roof displacements determined by linear dynamic analysis ($D_{th}$) and by modal response spectrum method ($D_{code}$)

## 4. CONCLUSIONS AND FINAL REMARKS

The efficacy of a recently proposed method (Borcia, 2010) of obtaining code-compliant sets of accelerograms was assessed. The method uses combined time and amplitude scaling of real seismic records, in conjunction with the requirement that all accelerograms in the set have a common, specified value of an instrumental Arias-type intensity.

The assessment was made with respect to the Romanian seismic code, P100-1/2006. As the provisions of this code are, to a large extent, close to those of Eurocode 8, some specifications in the European norm were used in parallel.

A set of ten accelerograms was obtained, by using the proposed method, corresponding to the elastic acceleration spectrum specified by the Romanian code for the city of Bucharest. A significant feature of this spectrum is the large corner period, $T_C$=1.6s.

Based on the results of the evaluations performed on SDOF and MDOF systems, it was concluded that the use of the considered set of accelerograms leads to a rather good estimation of the seismic action specified by the code.

**REFERENCES**

1. Borcia, I. S. (2010). „Recorded Accelerograms as an Alternative Description of the Seismic Action in the P 100-1/2006 Romanian Seismic Design Code". *Constructii*, Vol. 9, No. 2, pp. 79-88, http://constructii.incerc2004.ro/Archive/2010-2/Art6-2-2010.pdf (Last accessed: October 2012).

2. Borcia, I. S., Dobre, D. (2012). „Recorded accelerograms during strong Vrancea earthquakes and the P100-1/2006 Romanian seismic design code". *Proceedings of the 15th World Conference on Earthquake Engineering*, Lisbon, Portugal, September 24-28, Paper No. 1819.

3. CEN (2004). *Eurocode 8: Design of structures for earthquake resistance. Part 1: General rules, seismic actions and rules for buildings*. EN 1998-1:2004. Doc. CEN/TC250/SC8/N317. European Committee for Standardization.

4. Craifaleanu, I.-G. (2008). "A Comparison Between the Requirements of Present and Former Romanian Seismic Design Codes, Based on the Required Structural Overstrength". *Proceedings of the 14th World Conference of Earthquake Engineering*, Oct. 12-17, Beijing, China, Paper No. 08-01-0035 (on CD-ROM).

5. CSI (2009). *CSI Analysis Reference Manual for SAP2000, ETABS and SAFE*. Computers and Structures, Inc., Berkeley, California.

6. Grünthal, G. (Ed.) (1998). *European Macroseismic Scale 1998*. Cahiers du Centre Européen de Géodynamique et Séismologie, 15.

7. MTCT (2006). *P100-1/2006: Seismic Design Code. Part I. Design Rules for Buildings*. Construction Bulletin, No. 12-13, INCERC Bucharest Eds. (in Romanian).

8. Sandi, H. (1987). "Evaluation of seismic intensity based on instrumental data". *Constructii*, No. 6 - 7 (in Romanian).

9. Sandi, H., Floricel, I. (1998). "Some alternative instrumental measures of ground motion severity". *Proceedings of the 11th European Conference on Earthquake Engineering*, Paris, France, September 6-11.

10. Sandi, H. (Ed.), Aptikaev, F., Borcia, I. S., Erteleva, O., Alcaz, V. (2010). *Quantification of Seismic Action on Structures*. AGIR Publishing House, Bucharest, Romania.

11. Sandi H., Borcia I.S. (2011). "Intensity Spectra versus Response Spectra: Basic Concepts and Applications", *Pure and Applied Geophysics*, Vol. 168, Issue 1-2, pp. 261-287.